\documentclass[conference]{IEEEtran}
\IEEEoverridecommandlockouts
{ \large \normalsize}%
{ \large \normalsize}%

{ \large \normalsize}%
{ \large \normalsize}%

{ \large \normalsize}%
{ \large \normalsize}

\newtheorem{lemma}{Lemma}

\newtheorem{claim}[lemma]{Claim}

\newtheorem{theorem}[lemma]{Theorem}
\newtheorem{definition}[lemma]{Definition}

\newcommand{\nix}[1]{}

\def\QED{\mbox{\rule[0pt]{1.5ex}{1.5ex}}}

\newenvironment{proof}{\textit{Proof:}}{{\hspace*{\fill}~\QED\\}}

\input{psfig}
\usepackage{amsmath, amssymb}
\usepackage{color}
\newcommand{\calS}{{\cal S}} \newcommand{\calT}{{\cal T}}

\newcommand{\bfT}{\mathbf{T}}
\newcommand{\bfS}{\mathbf{S}}

\newcommand{\bfP}{\mathbf{P}}

\newcommand{\di}{d_i}
\newcommand{\dihat}{\hat{d}_i}
\newcommand{\uj}{u_j}
\newcommand{\ujhat}{\hat{u}_j}
\def\endproof{\hspace*{\fill}~$\blacksquare$}
\begin{document}

\title{Overlay Protection Against Link Failures Using Network Coding}
\author{ Ahmed E. Kamal, Aditya Ramamoorthy, Long Long, Shizheng Li
\thanks{The authors are with the Dept. of Electrical and Computer Engineering at Iowa State University, Ames, IA 50011 (email: \{kamal, adityar, longlong, szli\}@iastate.edu). The material in this paper has appeared in part at the 42nd Annual Conf. on Information Sciences and Systems (CISS), 2008. This work was funded in part by grants CNS-0626741 and CNS-0721453 from NSF, and a gift from Cisco Systems.}
}

\date{\today}
\maketitle
\begin{abstract}
This paper introduces a network coding-based protection scheme
against single and multiple link failures. The proposed strategy
ensures that in a connection, each node receives two copies of the
same data unit: one copy on the working circuit, and a second copy
that can be extracted from linear combinations of data units
transmitted on a shared protection path. This guarantees
instantaneous recovery of data units upon the failure of a working
circuit. The strategy can be implemented at an overlay layer,
which makes its deployment simple and scalable. While the proposed
strategy is similar in spirit to the work of Kamal '07 \& '10,
there are significant differences.
In particular, it provides protection against multiple link
failures. The new scheme is simpler, less expensive, and does not
require the synchronization required by the original scheme. The
sharing of the protection circuit by a number of connections is
the key to the reduction of the cost of protection. The paper also
conducts a comparison of the cost of the proposed scheme to the
1+1 and shared backup path protection (SBPP) strategies, and
establishes the benefits of our strategy.

\end{abstract}

\section{Introduction}
Research on techniques for providing protection to networks
against link and node failures has received significant attention
\cite{Zhou00}. Protection, which is a proactive technique, refers
to reserving backup resources in anticipation of failures, such
that when a failure takes place, the pre-provisioned backup
circuits are used to reroute the traffic affected by the failure.
Several protection techniques are well known, e.g., in 1+1 protection, the connection traffic is simultaneously transmitted on two link disjoint paths. The receiver, picks the path with the stronger signal. On the other hand in 1:1 protection, transmission on the backup path only takes place in the case of failure. Clearly, 1+1 protection provides instantaneous recovery from failure, at increased cost. 
However, the cost of protection circuits
is at least equal to the cost of the working circuits, and
typically exceeds it.
To reduce the cost of
protection circuits, 1:1 protection has been extended to 1:N
protection, in which one backup circuit is used to protect N
working circuits. However, failure detection and data rerouting are
still needed, which may slow down the recovery process. In order
to reduce the cost of protection, while still providing
instantaneous recovery, references \cite{Kamal06a, Kamal09} proposed
the sharing of one set of protection circuits by a number of
working circuits, such that each receiver in a connection is able
to receive two copies of the same data unit: one on the working
circuit, and another one from the protection circuit. Therefore,
when a working circuit fails, another copy is readily available
from the protection circuit. The sharing of the protection circuit
was implemented by transmitting data units such that they are
linearly combined inside the network, using the technique of
network coding \cite{Ahlswede00}. Two linear combinations are
formed and transmitted in two opposite directions on a p-Cycle
\cite{Grover04}. We refer to this technique as 1+N protection,
since one set of protection circuits is used to simultaneously
protect a number of working circuits. The technique was
generalized for protection against multiple failures in
\cite{Kamal07a}.

In this paper, we propose a new method for protection against
multiple failures that is related to the techniques of
\cite{Kamal09, Kamal07a}. Our overall objective is still the same;
however, the proposed scheme improves upon the previous techniques
in several aspects.
First, instead of cycles, we use paths to carry the
linear combinations. This reduces the cost of implementation even
further, since in the worst case the path
can be implemented using the cycle less one segment (that may consist of several links).
Moreover, a path may be feasible, while a cycle may not. Second,
each linear combination includes data units transmitted from the
same round, as opposed to transmitting data units from different
rounds as proposed in \cite{Kamal09}. This simplifies the
implementation and synchronization between nodes. This aspect is
especially important when considering a large number of protection
paths, since synchronization becomes a critical issue in this
case. The protocol implementation is therefore self-clocked since
data units at the heads of the local buffers in each node are
combined provided that they belong to the same round.
Overall, these improvements result in a simple and scalable protocol that can be implemented at the overlay layer.
The paper also includes details about implementing the proposed
strategy. A network coding scheme to protect against adversary
errors and failures under a similar model is proposed in
\cite{LiR2010}, in which more protection resources are required.

This paper is organized as follows. In Section \ref{sec:model} we
introduce our network model and assumptions. In Section
\ref{sec:single-fault} we introduce the modified technique for
protection against single failures. Implementation issues are
discussed in  Section \ref{sec:implementation}. In Section
\ref{sec:multiple-faults} we present a generalization of this
technique for protecting against multiple failures. The encoding
coefficient assignment is discussed in Section \ref{sec:coeff}. In
Section \ref{modified-formulation} we present an integer linear
programming formulation to provision paths to protect against
single failures. Section \ref{sec:results} provides some results
on the cost of implementing the proposed technique, and compares
it to 1+1 protection and SBPP. Section \ref{sec:conclusions}
concludes this paper with a few remarks.

\section{Model and Assumptions}
\label{sec:model}
 In this section we introduce our network model
and the operational assumptions. We also define a number of
variables and parameters which will be used throughout the paper.
\subsection{Network Model}
We assume that the network is represented by an undirected graph,
$G(V,E)$, where $V$ is the set of nodes and $E$ is the set of
edges. Each node corresponds to a switching node, e.g., a router,
a switch or a crossconnect. Network users access the network by
connecting to input ports of such nodes, possibly through
multiplexing devices. Each undirected edge corresponds to two
transmission links, e.g., fibers, which carry data in two opposite
directions. The capacity of each link is a multiple of a basic
transmission unit, which can be wavelengths, or smaller
tributaries, such as DS-3, or OC-3. In this paper, we do not
impose an upper limit on the capacity of a link, and we assume
that it carries a sufficiently large number of basic tributaries,
i.e., we consider the uncapacitated case.

In order to protect against single link failures, the network
graph needs to be at least 2-connected. That is, between each pair
of nodes, there needs to be at least two link disjoint paths. The
number of protection paths, and the connections protected by each
of these paths depends on the connections and their end points, as
well as the network graph. An example of connection protection in
NSFNET will be given in Section \ref{sec:single-fault}. In
general, for protection against $M$ link failures, the graph needs
to be $(M+1)$-connected.

Since providing protection to connections will require the use of
finite field arithmetic, these functions are better implemented in
the electronic domain. Therefore, we assume that protection is
provided at a layer that is above the optical layer, and this is
why we refer to this type of protection as \textit{overlay
protection}.

\subsection{Operational Assumptions}
\label{sec:assumptions}

We make the following operational assumptions:
\begin{enumerate}
\item
 The protection is at the connection level, and
it is assumed that all connections that are protected together
will have the same transport capacity, which is the maximum bit
rate that has to be handled by the connection. We refer to this
transport capacity as $B$\footnote{
Throughout this paper we assume that all connections that are
protected together have the same transport capacity. The case of
unequal transport capacities can also be handled, but will not be
addressed in this paper.}.
\item
All connections are bidirectional.
\item
Paths used by
connections that are jointly protected are link disjoint.
\item A set of
connections will be protected together by a protection path. The
protection path is bidirectional, and it passes through all end
nodes of the protected connections. The protection path is also
link disjoint from the paths used by the protected connections.
\item
 Links of the protection path protecting a set of
connections have the same capacity of these connections, i.e.,
$B$.
\item
 Segments of the protection path are terminated at each
connection end node on the path. The data received on the
protection path segment is processed, and retransmitted on the
outgoing port, except for the two extreme nodes on the protection
path.
\item Data
units are fixed and equal in size. \item Nodes are equipped with
sufficiently large buffers. The upper bound on buffer sizes will
be derived in Section \ref{sec:implementation}. 
\item When a link carrying active (working) circuits fails, the
receiving end of the link receives empty data units. We regard
this to be a data unit containing all zeroes.
\item
The system works in time slots. In each time slot a new data unit
is transmitted by each end node of a connection on its primary
path\footnote{The terms primary and working circuits, or paths,
will be used interchangeably.}.  In addition, this end node also
transmits a data unit in each direction on the protection path.
The exact specification of the protocol, and the data unit is
given later.
\item
The amount of time consumed in solving a system of equations is
negligible in comparison to the length of a time slot. This
ensures that the buffers are stable\footnote{
Typically, a single connection will have a bit rate on the order
of 10's or 100's of Mbps that is much lower than the capacity of a
fiber or a wavelength. Therefore, we assume that the processing
elements of a switching node will be able to process the data
units within the transmission time of one data unit. }.
\end{enumerate}

 The symbols used in this paper are listed in
Table \ref{tab:symbols}, and will be further explained within the
text. The upper half of the table defines symbols which relate to
the working, or primary connections, and the lower half introduces
the symbols used in the protection circuits.
\begin{table}
\caption{List of symbols: Upper half are symbols used for working
paths, and lower half are symbols for protection paths.}
\label{tab:symbols}
\begin{tabular}{l p{2.3in}}
\hline \hline
Symbol & Meaning\\
\hline
$\mathbb N$ & set of connections to be protected\\
$N$ & number of connections = $ \left| \mathbb N \right|$\\
$\calS$, $\calT$ & two disjoint ordered sets of communicating
nodes,
such that a node in $\calS$ communicates with a node in $\calT$\\
${\cal S}_k$, ${\cal T}_k$ & sets of connection end nodes protected by $\bfP_k$\\
$S_i$, $T_j$ & nodes in $\calS$ and $\calT$, respectively\\
$\di$, $\uj$ & data units sent by nodes $S_i$ and $T_j$, respectively\\
$\dihat$, $\ujhat$ & data units sent by nodes $S_i$ and $T_j$,
respectively, on the primary paths, which are received by their
respective receiver nodes\\
$T(S_i)$ & node in $\calT$ transmitting to and receiving from $S_i$\\
$S(T_j)$ & node in $\calS$ transmitting to and receiving from $T_j$\\
$B$ & the capacity protected by the protection path \\
$n$ & round number\\
\hline
$M$ & total number of failures to be protected against ($M=1$ in Section \ref{sec:single-fault}).\\
$\bfP$ (or $\bfP_k$) & bidirectional path used for protection\\
$\mathbb P$ & set of protection paths\\
$\bfS$, $\bfT$ & unidirectional paths of $\bfP$ started by $S_1$
and $T_1$, respectively\\
$\sigma(S_i)(\sigma(T_j))$& the next node downstream from $S_i$ (respectively $T_j$) on $\bfS$\\
$\sigma^{-1}(S_i)(\sigma^{-1}(T_j))$ & the next node upstream from $S_i$ (respectively $T_j$) on $\bfS$\\
$\tau(S_i)(\tau(T_j))$ & the next node downstream from $S_i$ (respectively $T_j$) on $\bfT$\\
$\tau^{-1}(S_i)(\tau^{-1}(T_j))$& the next node upstream from $S_i$ (respectively $T_j$) on $\bfT$\\
$\chi_w (\chi_{\bfP})$ & delay over working (protection) path\\
$F_\bfS (S_i) (F_\bfT (S_i))$ & buffers at node $S_i$ used for transmission on the $\bfS$ ($\bfT$) paths\\
$\alpha_{i \leftrightarrow j,k}$ & scaling coefficient used for connection between $S_i$ and $T_j$ on ${\bfP}_k$\\
$y_e (z_e)$ & The data unit transmitted on link $e\in \bfS$ (
$e\in \bfT$ respectively) \\
$K$ & The total number of protection paths, i.e., $|\mathbb{P}|$\\ \hline
\end{tabular}

\end{table}
All operations in this paper are over the finite field $GF(2^m)$
where $m$ is the length of the data unit in bits. It should be
noted that all addition operations (+) over $GF(2^m)$ can be
simply performed by bitwise XOR's. In fact, for protection against
single-link failures we only require addition operations, which
justifies the last assumption above.


\section{1+N Protection Against Single Link Failures}
\label{sec:single-fault}
In this section we introduce our strategy
for implementing network coding-based protection against single
link failures.


 Consider a set of $\mathbb{N}$ bidirectional,
unicast connections, where the number of connections is given by
$N=|\mathbb{N}|$.
Connection $i \leftrightarrow j$ is between nodes $S_i$ and $T_j$.
Nodes $S_i$ and $T_j$ belong to the two ordered sets $\calS$ and
$\calT$, respectively.
Data units are transmitted by nodes in $\calS$ and $\calT$ in
rounds, such that the data unit transmitted from $S_i$ to $T_j$ in
round $n$ is denoted by $d_i(n)$, and the data unit transmitted
from $T_j$ to $S_i$ in the same round is denoted by $u_j(n)$
\footnote{For simplicity, the round number, $n$, may be dropped
when it is obvious.}. The data units received by nodes $S_i$ and
$T_j$ are denoted by $\ujhat$ and $\dihat$, respectively, and can
be zero in the case of a failure on the primary circuit between
$S_i$ and $T_j$.

\begin{figure}
\centerline{\psfig{figure=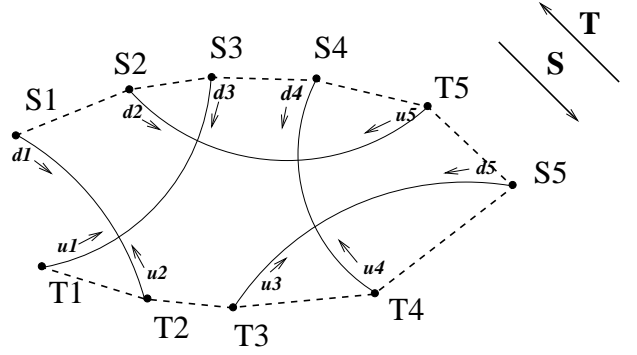,width=3.2in}}
\caption{An example of enumerating the nodes in five connections.
Node $T_5$ is the first node to be encountered while traversing
$\bfS$, which communicates with a node in $\calS$ that has already
been enumerated ($S_2$).} \label{fig:example-enumeration}
\end{figure}

The two ordered sets, $\calS = (S_1 , S_2 , \ldots , S_N)$ and
$\calT = (T_1 , T_2 , \ldots , T_N)$ are of equal lengths, $N$,
which is the number of connections that are jointly protected. If
two nodes communicate, then they must be in different ordered
sets. These two ordered sets define the order in which the
protection path, $\bfP$, traverses the connections' end nodes. The
ordered set of nodes in $\calS$ is enumerated in one direction,
and the ordered set of nodes in $\calT$ is enumerated in the
opposite direction on the path. The nodes are enumerated such that
one of the two end nodes of $\bfP$ is labeled $S_1$. Proceeding on
$\bfP$ and inspecting the next node, if the node does not
communicate with a node that has already been enumerated, it will
be the next node in $\calS$, using ascending indices for $S_i$.
Otherwise, it will be in $\calT$, using descending indices for
$T_i$. Therefore, node $T_1$ will always be the other end node on
$\bfP$. The example in Figure \ref{fig:example-enumeration} shows
how ten nodes, in five connections are assigned to $\calS$ and
$\calT$. The bidirectional protection path is shown as a dashed
line.

Under normal working conditions the working circuit will be used
to deliver $d_i$ and $u_j$ data units from $S_i$ to $T_j$ and from
$T_j$ to $S_i$, respectively. The basic idea for receiving a
second copy of data $u_j$ by node $S_i$, for example, is to
receive on two opposite directions on the protection path, ${\bf
P}$, the signals given by the following two equations, where all
data units belong to the same round, $n$:
\begin{eqnarray}
&&\sum_{k,~S_k \in \mathbf{A}} d_k  + \sum_{k,~T_k \in \mathbf{B}} \hat{u}_k \label{eqn:sum_left}\\
&&u_j + \sum_{k,~T_k\in \mathbf{B}} u_k + \sum_{k, ~S_k \in
\mathbf{A}} \hat{d}_k
  \label{eqn:sum_right}
\end{eqnarray}
where $\mathbf{A}$ and $\mathbf{B}$ are disjoint subsets of nodes
in the ordered set of nodes
 $\calS$ and $\calT$, respectively, such that a node in $\mathbf{A}$ communicates
with a node in $\mathbf{B}$, and vice versa. If the link between
$S_i$ and $T_j$ fails, then $u_j$ can be recovered by $S_i$
 by simply adding equations (\ref{eqn:sum_left}) and
(\ref{eqn:sum_right}).

We now outline the steps involved in the construction of the
primary/protection paths and the encoding/decoding operations at
the individual nodes.
\subsection{Protection Path Construction and Node Enumeration}

\begin{figure}
\psfig{figure=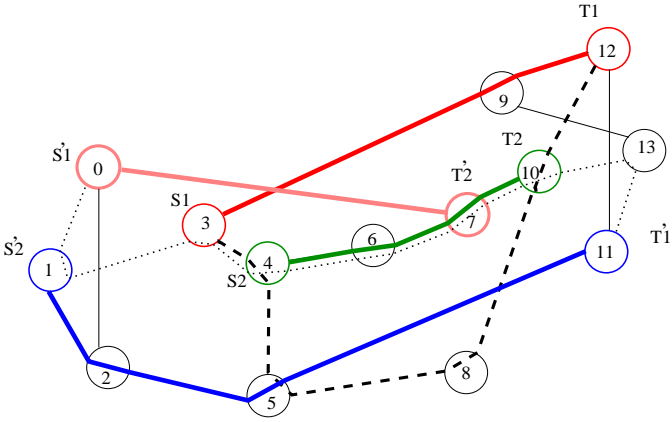,width=3.5in}
\caption{An example of provisioning and protecting four connections
on NSFNET.}
\label{fig:nsfnet-example}
\end{figure}

\begin{enumerate}
\item Find a bidirectional path\footnote{The path is not
necessarily a simple path, i.e., vertices and links may be
repeated. We make this assumption in order to allow the
implementation of our proposed scheme in networks where some nodes
have a nodal degree of two. Although the graph theoretic name for
this type of paths is a {\it walk}\/, we continue to use the term
{\it path}\/ for ease of notation and description.}, $\bfP$, that goes through all
the end nodes of the connections in $\mathbb N$. $\bfP$ consists
of two unidirectional paths in opposite directions.
These two unidirectional paths do not have to traverse the same
links, but must traverse the nodes in the opposite order. One of
these paths will be referred to as $\bfS$ and the other one as
$\bfT$.
\item
Given the set of nodes in all $N$ connections which are to be
protected together, construct the ordered sets of nodes, $\calS$
and $\calT$, as explained above
\typeout{ If two nodes communicate, then they must be in different
ordered sets. }
\typeout{ The ordered set of nodes in $\calS$ is enumerated in one
direction, and the ordered set of nodes in $\calT$ is enumerated
in the opposite direction on the path. The nodes are enumerated
such that one of the two end nodes of $\bfP$ is labeled $S_1$.
Proceeding on $\bfP$ and inspecting the next node, if the node
does not communicate with a node that has already been enumerated,
it will be the next node in $\calS$, using ascending indices for
$S_i$. Otherwise, it will be in $\calT$, using descending indices
for $T_i$. Therefore, node $T_1$ will always be the other end node
on $\bfP$. The example in Figure \ref{fig:example-enumeration}
shows how ten nodes, in five connections are assigned to $\calS$
and $\calT$. The bidirectional protection path is shown as a
dashed line. }
\item A node $S_i$ in $\calS$ ($T_j$ in $\calT$) transmits $d_i$
($u_j$) data units to a node in $\calT$ ($\calS$) on the primary path,
which is received
as $\dihat$ ($\ujhat$).
\item \label{item:half-cycles} Transmissions on the two
unidirectional paths $\bfS$ and $\bfT$ are in rounds, and are
started by nodes $S_1$ and $T_1$, respectively. All the processing of
data units occurs between data units belonging to the same round.
\end{enumerate}
It is to be noted that it may not be possible to protect all
connections together, and therefore it would be necessary to
partition the set of connections, and protect connections in each
partition together. We illustrate this point using the example
shown in Figure \ref{fig:nsfnet-example}, where there are four
connections (shown using bold lines) that are provisioned on
NSFNET: $C_1=(3,12)$, $C_2=(4,10)$, $C_3=(0,7)$ and $C_4=(1,11)$.
It is not possible to protect all four connections together using
one protection path that is link disjoint from all four
connections. Therefore, in this example, we use two protection
paths: one protection path (3,4,5,8,10,12) protecting $C_1$ and
$C_2$, and is shown in dashed lines; and another protection path
(0,1,3,4,6,7,10,13,11) protecting $C_3$ and $C_4$, and is shown in
dotted lines. Notice that all connections that are protected
together, and their protection path are link disjoint. The end
nodes in $C_1$ and $C_2$ are labeled $S_1$, $S_2$, $T_1$ and
$T_2$, while the end nodes in $C_3$ and $C_4$ are labeled
$S^\prime_1$, $S^\prime_2$, $T^\prime_1$ and $T^\prime_2$,
respectively.
In the above example, it is assumed that each connection is
established at an electronic layer, i.e., an overlay layer above
the physical layer. For example, the working path of a connection
can be routed and established as an MPLS Label Switched Path
(LSP), which can be explicitly routed in the network, as shown in
the figure, and therefore the paths of the connections which are
jointly protected, e.g., $C_1$ and $C_2$ in the above example, can
be made link disjoint. However, when it comes to the protection
path, since the data units transmitted on this path need to be
processed, the protection path can be provisioned as segments,
where each segment is an MPLS LSP which is explicitly routed. For
the example of Figure \ref{fig:nsfnet-example}, the protection
path protecting connections $C_1$ and $C_2$ can be provisioned as
three MPLS LSPs, namely, (3,4), (4,5,8,10) and (10,12).

\subsection{Encoding Operations on $\bfS$ and $\bfT$}
\label{sec:encoding}

The network encoding operation is executed by each node in $\calS$
and $\calT$. To facilitate the specification of the encoding
protocol we first define the following.
\begin{itemize}
\item $T(S_i)$: node in $\calT$ transmitting to and receiving from
$S_i$, e.g. in Fig.1, $T(S_1) = T_2$. \item $S(T_j)$: node in
$\calS$ transmitting to and receiving from $T_j$. \item
$\sigma(S_i)/\sigma(T_j)$: the next node downstream from $S_i$
(respectively $T_j$) on $\bfS$, e.g., in Fig.1, $\sigma(S_2) =
S_3$. \item $\sigma^{-1}(S_i)/\sigma^{-1}(T_j)$: the next node
upstream from $S_i$ (respectively $T_j$) on $\bfS$, e.g., in
Fig.1, $\sigma^{-1}(T_5) = S_4$. \item $\tau(S_i)/\tau(T_j)$: the
next node downstream from $S_i$ (respectively $T_j$) on $\bfT$,
e.g., in Fig. 1, $\tau(T_4) = S_5$. \item
$\tau^{-1}(S_i)/\tau^{-1}(T_j)$: the next node upstream from $S_i$
(respectively $T_j$) on $\bfT$,e.g., in Fig.1, $\tau^{-1}(S_5) =
T_4$.
\end{itemize}
We denote the data unit transmitted on link $e \in \bfS$ by $y_e$
and the data unit transmitted on link $e \in \bfT$ by $z_e$.
Assume that nodes $S_i$ and $T_j$ are in the same connection. The
encoding operations work as follows,  where all data units belong
to the same round.

\begin{enumerate}
\item {\it Encoding operations at $S_i$.} The node $S_i$ has
access to data units $d_i$ (that it generated) and data unit
$\ujhat$ received on the primary path from $T_j$.
\begin{enumerate}
\item It computes $y_{\sigma^{-1}(S_i) \rightarrow S_i} + (d_i +
\ujhat)$ and sends it on the link $S_i \rightarrow \sigma(S_i)$;
i.e.
\begin{align*}
y_{S_i \rightarrow \sigma(S_i)} &= y_{\sigma^{-1}(S_i) \rightarrow
S_i} + (d_i + \ujhat).
\end{align*}
\item It computes $z_{\tau^{-1}(S_i) \rightarrow S_i} + (d_i +
\ujhat)$ and sends it on the link $S_i \rightarrow \tau(S_i)$;
i.e.
\begin{align*}
z_{S_i \rightarrow \tau(S_i)} &= z_{\tau^{-1}(S_i) \rightarrow
S_i} + (d_i + \ujhat).
\end{align*}
\end{enumerate}

\item {\it Encoding operations at $T_j$.} The node $T_j$ has
access to data units $u_j$ (that it generated) and data unit
$\dihat$ received on the primary path from $S_i$.
\begin{enumerate}
\item It computes $y_{\sigma^{-1}(T_j) \rightarrow T_j} + (\dihat +
u_j)$ and sends it on the link $T_j \rightarrow \sigma(T_j)$; i.e.

\begin{align*}
y_{T_j \rightarrow \sigma(T_j)} &= y_{\sigma^{-1}(T_j) \rightarrow
T_j} + (\dihat + u_j)
\end{align*}
\item It computes $z_{\tau^{-1}(T_j) \rightarrow T_j} + (\dihat
+ u_j)$ and sends it on the link $T_j \rightarrow \tau(T_j)$;
i.e.
\begin{align*}
z_{T_j \rightarrow \tau(T_j)} &= z_{\tau^{-1}(T_j) \rightarrow
T_j} + (\dihat + u_j)
\end{align*}
\end{enumerate}
\end{enumerate}

\begin{figure}
\centerline{\psfig{figure=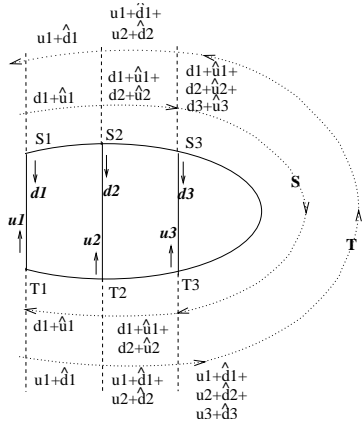,width=2.2in}}
\caption{Example of three nodes performing the encoding
procedure.
Note that the addition (bitwise XOR) of two copies of the same data unit, e.g.,
$d_i$ and $\hat{d}_i$, removes both of them.}
\label{fig:example-one-failure}
\end{figure}

An example in which three nodes perform this procedure in the absence
of failures is shown in Figure \ref{fig:example-one-failure}.

Consider $S' \subseteq \calS$ and let $\mathcal{N}(S')$ represent
the subset of nodes in $\calT$ that have a primary path connection
to the nodes in $S'$ (similar notation shall be used for a subset
$T' \subseteq \calT$). Let $D_{\bfS}(S_i)$ and $U_{\bfS}(S_i)$
represent the set of downstream and upstream nodes of $S_i$ on the
protection path $\bfS$ (similar notation shall be used for the
protection path $\bfT$). When all nodes in $\calS$ and $\calT$
have performed their encoding operations, the signals received at
a node $S_i$ on the $\bfS$ and $\bfT$ paths, respectively, are as
follows
\begin{align}
&y_{\sigma^{-1}(S_i) \rightarrow S_i} \nonumber\\
&= \underbrace{\sum_{\{k: S_k \in U_{\bfS}(S_i) \cap \calS\}} d_k
+ \sum_{\{k: T_k \in \mathcal{N}(U_{\bfS}(S_i) \cap
\calS)\}} \hat{u}_k}_{\text{From nodes upstream of $S_i$ on $\bfS$ in~} \calS} \nonumber\\
&+\underbrace{\sum_{\{k: T_k \in U_{\bfS}(S_i) \cap \calT\}} u_k +
\sum_{\{k: S_k \in \mathcal{N}(U_{\bfS}(S_i) \cap \calT)\}}
\hat{d}_k}_{\text{From nodes upstream of $S_i$ on $\bfS$ in~}
\calT} , \text{~and} \label{eqn:val-from-S} \\
&z_{\tau^{-1}(S_i) \rightarrow S_i} \nonumber\\&=
\underbrace{\sum_{\{k: S_k \in U_{\bfT}(S_i) \cap \calS\}} d_k +
\sum_{\{k: T_k \in
\mathcal{N}(U_{\bfT}(S_i) \cap \calS)\}} \hat{u}_k}_{\text{From nodes upstream of $S_i$ on $\bfT$ in~} \calS} \nonumber\\
&+\underbrace{\sum_{\{k: T_k \in U_{\bfT}(S_i) \cap \calT\}} u_k +
\sum_{\{k: S_k \in \mathcal{N}(U_{\bfT}(S_i) \cap \calT)\}}
\hat{d}_k}_{\text{From nodes upstream of $S_i$ on $\bfT$ in~}
\calT} \label{eqn:val-from-T}
\end{align}
Similar equations can be derived for node $T_j$.\\
\subsection{Recovery from failures}
\label{sec:single-failure-recovery}

The encoding operations described in Subsection \ref{sec:encoding}
allow the recovery of a second copy of the same data unit
transmitted on the working circuit, hence protecting against
single link failures. To illustrate this, suppose that the primary
path between nodes $S_i$ and $T_j$ fails. In this case, $S_i$ does
not receive $u_j$ on the primary path, and it receives $\ujhat =
0$ instead. Moreover, $\dihat = 0$. However, $S_i$ can recover
$u_j$ by adding equations (\ref{eqn:val-from-S}) and
(\ref{eqn:val-from-T}). In particular node $S_i$ computes
\begin{align}
y_{\sigma^{-1}(S_i) \rightarrow S_i} + z_{\tau^{-1}(S_i)
\rightarrow S_i} &= \sum_{\{k: S_k \in \calS \backslash \{S_i\}\}}
d_k +
\sum_{\{k: T_k \in \calT\}} u_k \nonumber\\
&+ \sum_{\{k: T_k \in \calT \backslash \{T_j\} \}} \hat{u}_k +
\sum_{\{k: S_k \in \calS\}} \hat{d}_k \nonumber\\
&= \hat{d}_i + u_j \nonumber\\
&= u_j \text{~~(since $\hat{d}_i = 0$.)}
\end{align}

Similarly, $T_j$ can recover $d_i$ by adding the values it obtains
over $\bfS$ and $\bfT$ . For example,
if the working path between $S_2$ and $T_2$ in Figure
\ref{fig:example-one-failure} fails,
then at node $S_2$ adding the signal received on
\textbf{S} to the signal received on \textbf{T},
then $u_2$ can be recovered, since $T_2$ generated $u_2$. Also,
node $T_2$ adds the signals on \textbf{S} and \textbf{T}
to recover $d_2$.

Notice that the reception of a second copy of $u_2$ and $d_2$ at
$S_2$ and $T_2$, respectively, when there are no failures,
requires the addition of the $d_2$ and $u_2$ signals generated by
the same nodes, respectively.

As a more general example, consider the case in Figure
\ref{fig:example-enumeration}. Node $S_5$, for example, will
receive the following signal on $\bfS$:
\begin{equation}
(d_1 + \hat{u}_2) + (d_2 + \hat{u}_5) + (d_3 + \hat{u}_1) + (d_4 +
\hat{u}_4) + (u_5 + \hat{d}_2), \label{eqn:S5-S}
\end{equation}
and will receive the following on $\bfT$:
\begin{equation}
(u_1 + \hat{d}_3) + (u_2 + \hat{d}_1) + (u_3 + \hat{d}_5) + (u_4 +
\hat{d}_4). \label{eqn:S5-T}
\end{equation}
If the link between $S_5$ and $T_3$ fails, then $\hat{d}_5 = 0$,
and adding equations (\ref{eqn:S5-S}) and (\ref{eqn:S5-T}) will
recover $u_3$ at $S_5$.


\section{Implementation Issues}
\label{sec:implementation}

In this subsection we address a number of practical implementation
issues.
\subsection{Round Numbers}
Since linear combinations include packets belonging to the same
round number, the packet header should include a round number
field. The field is initially reset to zero, and is updated
independently by each node when it generates and sends a new
packet on the working circuit. Note that there will be a delay
before the linear combination propagating on $\bfS$ and $\bfT$
reaches a given node. For example, in Figure
\ref{fig:example-one-failure} assuming that all nodes started
transmission at time $0$, node $S_3$ shall receive the combination
corresponding to round $0$ over $\bfS, d_1(0) + \hat{u}_1(0) +
d_2(0) + \hat{u}_2(0)$ after a delay corresponding to the
propagation delay between nodes $S_1$ and $S_3$, in addition to
the processing and transmission times at nodes $S_1$ and $S_2$.
However since the received data unit shall contain the round
number $0$, it shall be combined with the data unit generated by
$S_3$ at time slot $0$.


The size of the round number field depends on the delay of the
protection path, including processing and transmission times, as
well as propagation time, and the working circuit delay. It is
reasonable to assume that the delay of any working circuit is
shorter than that of the protection circuit; otherwise, the
protection path could have been used as a working path. Thus, when
a data unit on the protection path corresponding to a particular
round number reaches a given node, the data unit of that round
number would have already been received on the primary path of the
node.

 In this
case, it is straightforward to see that once a data unit is
transmitted on the working circuit, then it will take no more than
twice the delay of the protection path to recover the backup copy
of this data unit by the receiver. Therefore, round numbers can
then be reused. Based on this argument, the size of the set of
required unique round numbers is upper bounded by $2a$, where
\begin{equation}
a = \lceil \frac{\chi_{\bf
P}}{(Protection~data~unit~size~in~bits)/B} \rceil ~~.
\label{eqn:a}
\end{equation}
$\chi_{\bf P}$ in the above equation is the delay over
the protection circuit, and $B$ is the transport
capacity of the protection circuit, which, as stated in Section \ref{sec:assumptions},
is taken as the maximum over all the transport capacities of the protected connections.
A sufficiently long round number field will require no
more than $\log_2 (2a)$ bits.

\subsection{Synchronization}

An important issue is node synchronization to rounds. This can be
achieved using a number of strategies. A simple strategy for
initialization and synchronization is the following:
\begin{itemize}
\item In addition to buffers used to store transmitted and
received data units, each node $S_i \in \calS$ has two buffers,
$F_{\bf S} (S_i) $ and $F_{\bf T} (S_i)$, which are used for
transmissions on the $\bfS$ and $\bfT$ paths, respectively. Node
$T_j \in \calT$ also has similar buffers, $F_{\bf S} (T_j) $ and
$F_{\bf T }(T_j)$.
\item Node $S_1$ starts the transmission of $d_1 (0)$ on the
working circuit to $T(S_1)$. When $S_1$ receives $\hat{u}_{T(S_1)}
(0)$, it forms $d_1 (0) + \hat{u}_{T(S_1)} (0)$ and transmits it
on the outgoing link in $\bfS$. Similarly, node $T_1$ will
transmit $u_1 (0)$ on the working circuit, and  $u_1 (0) +
\hat{d}_{S(T_1)} (0)$ on the outgoing link in $\bfT$. \item Node
$S_i$, for $i > 0$, will buffer the combinations received on
$\bfS$ in $F_{\bf S} (S_i)$. Assume that the combination with the
smallest round number buffered in $F_{\bf S} (S_i)$ (i.e., head of
buffer) corresponds to round number $n$. When $S_i$ transmits $d_i
(n)$ and receives $\hat{u}_{T(S_i)} (n)$, then it adds those data
units to the combination with the smallest round number in $F_{\bf
S} (S_i)$ and transmits the combination on $\bfS$. The combination
with round number $n$ is then purged from $F_{\bf S} (S_i)$.
Similar operations are performed on $F_{\bf T} (S_i)$, $F_{\bf S}
(T_j)$ and $F_{\bf T} (T_j)$. Note that purging of the data unit
from the buffer only implies that the combination corresponding to
round $n$ has been sent and should not be sent again. However node
$S_i$ needs to ensure that it saves the value of the data unit
received on $\bfS$ as long as needed for it to be able to decode
${u}_{T(S_i)} (n)$ if needed. An illustration of the use of those
buffers is shown in Figure \ref{fig:node}.
\begin{figure}
\centerline{\psfig{figure=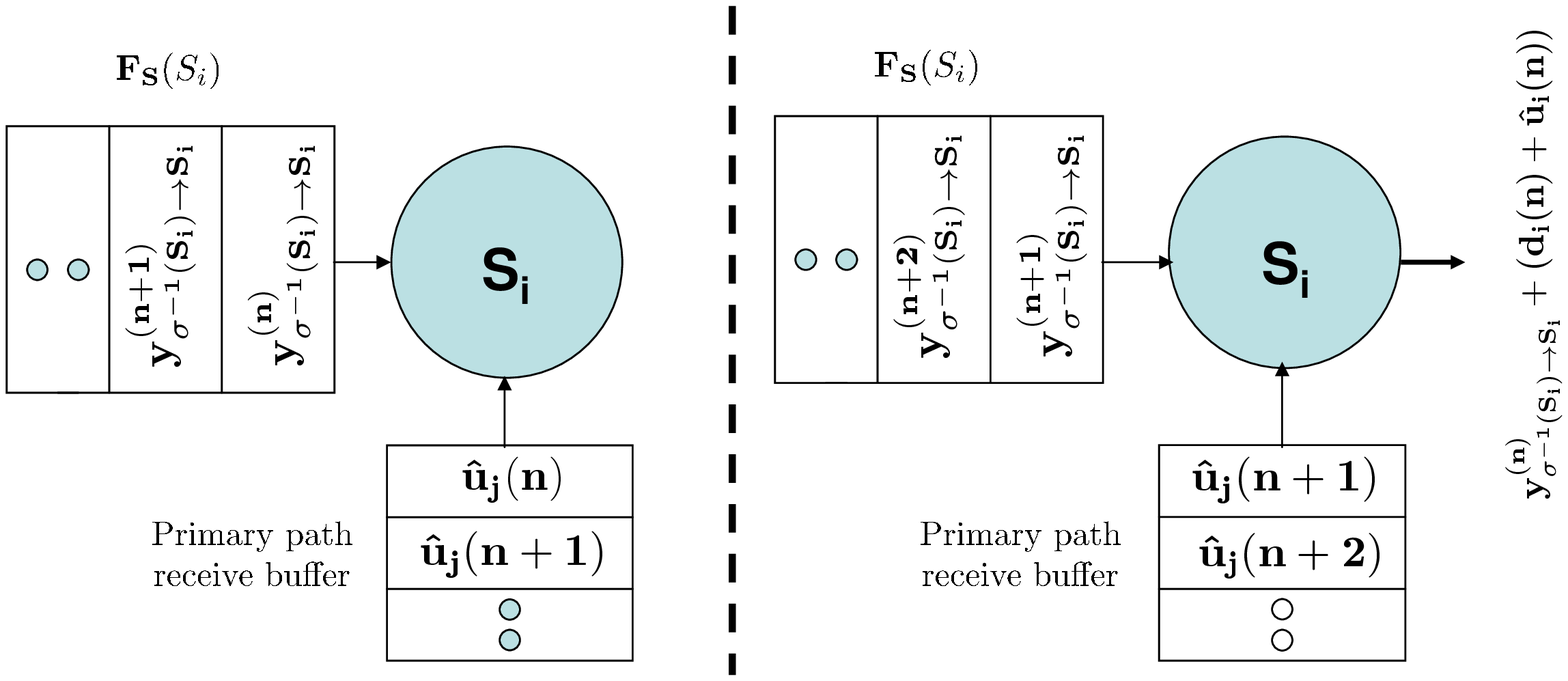,width=3.25in}}
\caption{An illustration of the use of node buffer $F_{\bf S}
(S_i)$. (a) Shows the status of the buffers before data unit at
round $n$ has been processed. (b) Shows the status of the buffers
after the data unit at round $n$ has been processed. Note that the
data units corresponding to round $n$ have been purged from both
$F_{\bf S} (S_i)$ and the primary path receive buffer. The
operation of other buffers is similar.} \label{fig:node}
\end{figure}

\end{itemize}

\subsection{Buffer Size}
Assuming that all nodes start transmitting simultaneously, then
all nodes would have decoded the data units corresponding to a
given round number in a time that does not exceed
\[
\chi_{\bf P} + \max_{1 \leq w \leq N} \chi_w
\]
where $\chi_w$ is the delay over working path $w$.

Based on this, the following upper bounds on buffer sizes can be
established:
\begin{itemize}
\item The transmit buffer, as well as the $F_{\bf S}$ and $F_{\bf T}$ buffers
are upper bounded by
\[
\lceil \frac{\chi_{\bf P} + \max_{1 \leq w \leq N}
  \chi_w}{Data~unit~size~in~bits/B} \rceil ~~.
\]
This is because it will take $\chi_w$ units of time over the path
$w$ used by the connection $S(T_1) \leftrightarrow T_1$ to receive
$\hat{d}_{S(T_1)}$, and then start
transmission on the $\bfT$ path.
An additional $\chi_P$ units
of time is required for the first combination to reach $S_1$. The numerator in
the above equation is the maximum of this delay. \item The receive
buffer is upper bounded by
\[
\lceil \frac{\chi_{\bf P} + \max_{1 \leq w \leq N} \chi_w -
\min_{1
    \leq w \leq N} \chi_w} {Data~unit~size~in~bits/B} \rceil ~~.
\]
The
numerator in the above equation is derived using arguments similar to the
transmit buffer, except that for the first data unit to be received, it will
have to encounter the delay over the working circuit; hence, the subtraction of
the minimum such delay.
\end{itemize}

\section{Protection against multiple faults}
\label{sec:multiple-faults}
We now consider the situation when
protection against multiple (more than one) link failures is
required. In this case it is intuitively clear that a given
primary path connection needs to be protected by multiple
bi-directional protection paths. To see this we first analyze the
sum of the signals received on $\bfS$ and $\bfT$ for a node $S_i$
that has a connection to node $T_j$ when the primary paths $S_i
\leftrightarrow T_j$ and $S_{i'} \leftrightarrow T_{j'}$ protected
by the same protection path are in
failure. In this case we have $\hat{d}_i = \hat{d}_{i'} =
\hat{u}_j = \hat{u}_{j'} = 0$. Therefore, at node $S_i$ we have,
\begin{align*}
y_{\sigma^{-1}(S_i) \rightarrow S_i} + z_{\tau^{-1}(S_i)
\rightarrow S_i} &= \sum_{\{k: S_k \in \calS \backslash \{S_i\}\}}
d_k +
\sum_{\{k: T_k \in \calT\}} u_k \nonumber\\
&+ \sum_{\{k: T_k \in \calT \backslash \{T_j\} \}} \hat{u}_k +
\sum_{\{k: S_k \in \calS\}} \hat{d}_k \nonumber\\
&= (d_{i'} + u_{j'}) + u_j.
\end{align*}
Note that node $S_i$ is only interested in the data unit $u_j$ but
it can only recover the sum of $u_j$ and the term $(d_{i'} +
u_{j'})$, in which it is not interested.

We now demonstrate that if a given connection is protected by
multiple protection paths, a modification of the protocol
presented in Section \ref{sec:encoding} can enable the nodes to
recover from multiple failures. In the modified protocol a node
multiplies the sum of its own data unit and the data unit received
over its primary path by an appropriately chosen scaling
coefficient before adding it to the signals on the protection
path. The scheme in Section \ref{sec:encoding} can be considered
to be a special case of this protocol when the scaling coefficient
is $1$ (i.e., the identity element over $GF(2^m)$).

It is important to note that in contrast to the approach presented
in \cite{Kamal07a}, this protocol does not require any
synchronization between the operation of the different protection
paths.

As before, suppose that there are $N$ bi-directional unicast
connections that are to be protected against the failure of any
$M$ links, for $M \leq N$. These connections are now protected by
$K$ protection paths $\bfP_k, k = 1, \dots, K$. Protection path
$\bfP_k$ passes through all nodes $\calS_k \subseteq \calS$ and
$\calT_k \subseteq \calT$ where the nodes in $\calS_k$ communicate
bi-directionally with the nodes in $\calT_k$. Note that
$\cup_{k=1}^K \calS_k = \calS$ and $\cup_{k=1}^K \calT_k = \calT$.
The ordered sets $\calS_k$ and $\calS_l$ are not necessarily
disjoint for $l \neq k$, i.e., a primary path can be protected by
different protection paths. However, if two protection paths are
used to protect the same working connection, then they must be
link disjoint.

\subsection{Modified Encoding Operation} \label{sec:mult_fault_encoding}
Assume that nodes $S_i$ and $T_j$ are protected by the protection
path $\bfP_k$. The encoding operations performed by $S_i$ and
$T_j$ for path $\bfP_k$ are explained below (the operations for
other protection paths are similar). In the presentation below we
shall use the notation $\sigma(S_i), \sigma^{-1}(S_i), \tau(S_i),
\tau^{-1}(S_i)$ to be defined implicitly over the protection
path $\bfP_k$.
Similar notation is used for $T_j$.

The nodes $S_i$ and $T_j$ initially agree on a value of the scaling
coefficient denoted $\alpha_{i \leftrightarrow j, k} \in GF(2^m)$.
The subscript $i\leftrightarrow j, k$ denotes that
the scaling coefficient is used for connection $S_i$ to $T_j$ over
protection path $\bfP_k$.
\begin{enumerate}
\item {\it Encoding operations at $S_i$.} The node $S_i$ has
access to data units $d_i$ (that it generated) and data unit
$\ujhat$ received on the primary path from $T_j$.
\begin{enumerate}
\item It computes $y_{\sigma^{-1}(S_i) \rightarrow S_i} + \alpha_{i
\leftrightarrow j, k} (d_i + \ujhat)$ and sends it on the link
$S_i \rightarrow \sigma(S_i)$; i.e.
\begin{align*}
y_{S_i \rightarrow \sigma(S_i)} &= y_{\sigma^{-1}(S_i) \rightarrow
S_i} + \alpha_{i \leftrightarrow j, k} (d_i + \ujhat).
\end{align*}
\item It computes $z_{\tau^{-1}(S_i) \rightarrow S_i} +
\alpha_{i \leftrightarrow j, k} (d_i + \ujhat)$ and sends it on
the link $S_i \rightarrow \tau(S_i)$; i.e.
\begin{align*}
z_{S_i \rightarrow \tau(S_i)} &= z_{\tau^{-1}(S_i) \rightarrow
S_i} +\alpha_{i \leftrightarrow j, k} (d_i + \ujhat).
\end{align*}
\end{enumerate}

\item {\it Encoding operations at $T_j$.} The node $T_j$ has
access to data units $u_j$ (that it generated) and data unit
$\dihat$ received on the primary path from $S_i$.
\begin{enumerate}
\item It computes $y_{\sigma^{-1}(T_j) \rightarrow T_j} +
\alpha_{i \leftrightarrow j, k} (\dihat + u_j)$ and sends it on
the link $T_j \rightarrow \sigma(T_j)$; i.e.
\begin{align*}
y_{T_j \rightarrow \sigma(T_j)} &= y_{\sigma^{-1}(T_j) \rightarrow
T_j} + \alpha_{i \leftrightarrow j, k} (\dihat + u_j)
\end{align*}
\item It computes $z_{\tau^{-1}(T_j) \rightarrow T_j} + \alpha_{i
\leftrightarrow j, k} (\dihat + u_j)$ and sends it on the link
$T_j \rightarrow \tau(T_j)$; i.e.
\begin{align*}
z_{T_j \rightarrow \tau(T_j)} &= z_{\tau^{-1}(T_j) \rightarrow
T_j} + \alpha_{i \leftrightarrow j, k}(\dihat + u_j)
\end{align*}
\end{enumerate}
\end{enumerate}
It should be clear that we can find expressions similar to the
ones in (\ref{eqn:val-from-S}) and (\ref{eqn:val-from-T}) in this
case as well.

\subsection{Recovery from failures}
\label{sec:reco_mul_fail} Suppose that the primary paths $S_i
\leftrightarrow T_j$ and $S_{i'} \leftrightarrow T_{j'}$ fail, and
they are both protected by $\bfP_k$. Consider the sum of the
signals received by node $S_i$ over $\bfS_k$ and $\bfT_k$. Similar
to our discussion in \ref{sec:single-failure-recovery}, we can
observe that
\begin{align*}
y_{\sigma^{-1}(S_i) \rightarrow S_i} + z_{\tau^{-1}(S_i)
\rightarrow S_i} &= \alpha_{i' \leftrightarrow j', k} (d_{i'} +
u_{j'}) + \alpha_{i \leftrightarrow j, k} u_j
\end{align*}
Note that the structure of the equation allows the node $S_i$ to
treat $(d_{i'} + u_{i'})$ as a single unknown. Thus from
protection path $\bfP_k$, node $S_i$ obtains one equation in two
variables. Now, if there exists another protection path $\bfP_{l}$
that also protects the connections $S_i \leftrightarrow T_j$ and
$S_{i'} \leftrightarrow T_{j'}$, then we can obtain the following
system of equations in two variables
\begin{align}
\label{eqn:system-of-eqs}
\begin{bmatrix}
\alpha_{i' \leftrightarrow j', k} & \alpha_{i \leftrightarrow j,
k}\\
\alpha_{i' \leftrightarrow j', l} & \alpha_{i \leftrightarrow j,
l}
\end{bmatrix}
\begin{bmatrix}
(d_{i'} + u_{j'})\\
u_j
\end{bmatrix}
= \begin{bmatrix} x^k_{S_i} \\ x^{l}_{S_i}\end{bmatrix},
\end{align}
where $x^k_{S_i}$ and $x^{l}_{S_i}$ represent values that can be
obtained at $S_i$ and therefore $u_j$ can be recovered by solving
the system of equations. The choice of the scaling coefficients
needs to be such that the associated $2 \times 2$ matrix in
(\ref{eqn:system-of-eqs}) is invertible. This can be guaranteed by
a careful assignment of the scaling coefficients. More generally
we shall need to ensure that a large number of such matrices need
to be full-rank. By choosing the operating field size $GF(2^m)$ to
be large enough, i.e., $m$ to be large enough we can ensure that
such an assignment of scaling coefficients always exists
\cite{harvey05}. The detailed discussion of coefficient
assignment can be found in Section \ref{sec:coeff}.

\subsection{Conditions for Data Recovery:}
\label{sec:conditions}

We shall first discuss the conditions for data recovery under a
certain failure pattern. To facilitate the discussion on
determining which failures can be recovered from, we represent the
failed connections, and the protection paths using a bipartite
graph, $G_{DR}(V, E)$, where the set of vertices $V = \mathbb N
\cup \mathbb P$, and the set of edges $E \subseteq \mathbb N
\times \mathbb P$ where $\mathbb N$ is the set of connections to
be protected, and $\mathbb P$ is the set of protection paths.
There is an edge from connection $N_i \in \mathbb N$ to protection
path $\bfP_k \in \mathbb P$ if $\bfP_k$ protects connection $N_i$.
In addition, each edge has a label that is assigned as follows.
Suppose that there exists an edge between $N_i$ (between nodes
$S_{i'}$ and $T_{j'}$) and $\bfP_k$. The label on the edge is
given by the scaling coefficient $\alpha_{i' \leftrightarrow j',
k}$.

Note that in general one could have link failures on primary paths
as well as protection paths. Suppose that a failure pattern is
specified as a set $F = \{N_{i_1}, \dots N_{i_n}\} \cup
\{\bfP_{j_1}, \dots, \bfP_{j_{n'}}\}$ where $\{N_{i_1}, \dots
N_{i_n}\}$ denotes the set of primary paths that have failed and
$\{\bfP_{j_1}, \dots, \bfP_{j_{n'}}\}$ denotes the set of
protection paths that have failed. The determination of whether a
given node can recover from the failures in $F$ can be performed
in the following manner.
\begin{figure}[ht]
\centerline{\psfig{figure=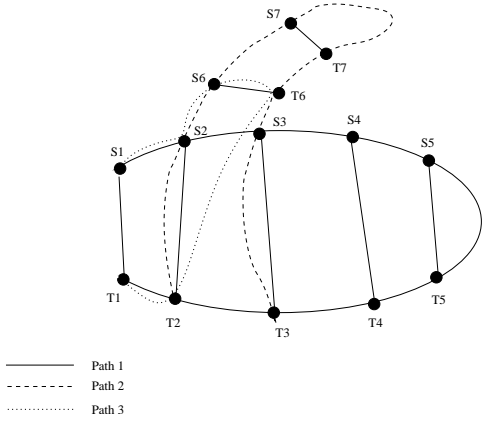,width=2.5in}}
\caption{An example of a network protected against multiple
faults.} \label{fig:3-paths-example}
\end{figure}

\begin{enumerate}
\item {\it Initialization.} Form the graph $G_{DR}(V, E)$ as
explained above. \item {\it Edge pruning.}
\begin{enumerate}
\item For all connections $N_i \in \mathbb{N} \setminus F$ remove
$N_i$ and all edges in which it participates from $G_{DR}$. \item
For all protection paths $\bfP_i \in F$ remove $\bfP_i$ and all
edges in which it participates from $G_{DR}$.
\end{enumerate}
\item {\it Checking the system of equations.} Let the residual
graph be denoted $G_{DR}^{'} = (\mathbb{N}^{'} \cup
\mathbb{P}^{'}, \mathbb{E}^{'})$. For each connection $N_i \in
\mathbb{N}^{'}$, do the following steps.
\begin{enumerate}
\item Let the subset of nodes in $\mathbb{P}^{'}$ that have a
connection to $N_i$ be denoted $\mathcal{N}(N_i)$. Each node in
$\mathcal{N}(N_i)$ corresponds to a linear equation that is
available to the nodes participating in $N_i$. The linear
combination coefficients are determined by the labels of the
edges.
Identify this system of equations. \item Check to see
whether a node in $N_i$ can solve this system of equations to
obtain the data unit it is interested in.
\end{enumerate}
\end{enumerate}

\begin{figure}
\centerline{\psfig{figure=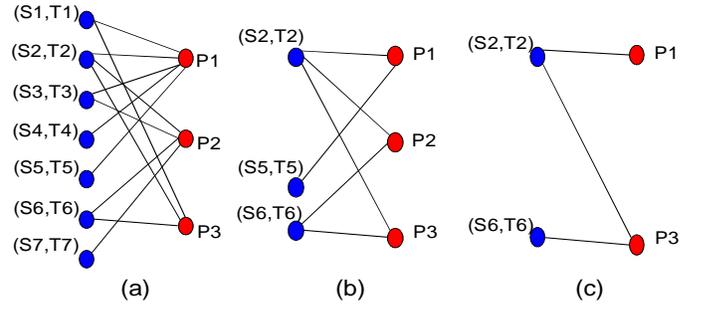,width=3.5in}}
\caption{Applying the bipartite graph representation verify if
  failures will be recovered.}
\label{fig:bipartite}
\end{figure}

 In Figure \ref{fig:bipartite} we show an example
that applies to the network in Figure \ref{fig:3-paths-example}.
Figure \ref{fig:bipartite}.(a) shows the bipartite graph for the
entire network, while Figures \ref{fig:bipartite}.(b) and
\ref{fig:bipartite}.(c) show the graph corresponding to the
following two failing patterns, respectively:
\begin{itemize}
\item ($S_2,T_2$), ($S_6,T_6$) and ($S_5,T_5$)
\item $\bfP_2$, ($S_2,T_2$) and ($S_6,T_6$)
\end{itemize}
 Let us assume that the encoding coefficients
are chosen to make sure the equation obtained by each node has
unique solution. From Figure \ref{fig:bipartite}.(b), the failures
of connections ($S_2,T_2$) and ($S_6,T_6$) can be recovered from
because each node obtains two equations in two unknowns. More
specifically, at node $S_2$ we obtain the following system of
equations (the equation from $\bfP_1$ is not used).
\begin{align*}
\begin{bmatrix}
\alpha_{2 \leftrightarrow 2, 2} & \alpha_{6 \leftrightarrow 6,
2}\\
\alpha_{2 \leftrightarrow 2, 3} & \alpha_{6 \leftrightarrow 6, 3}
\end{bmatrix}
\begin{bmatrix}
u_2\\
(d_6 + u_6)
\end{bmatrix}
= \begin{bmatrix} x^2_{S_2} \\ x^{3}_{S_2}\end{bmatrix},
\end{align*}
which has a unique solution if $(\alpha_{2 \leftrightarrow 2, 2}
\alpha_{6 \leftrightarrow 6, 3} - \alpha_{2 \leftrightarrow 2,
3}\alpha_{6 \leftrightarrow 6, 2}) \neq 0$. As pointed out in
Section \ref{sec:reco_mul_fail}, the choice of the scaling
coefficients can be made so that all possible matrices involved
have full rank by working over a large enough field size. Thus in
this case $S_2$ and $T_2$ can recover from the failures. By a
similar argument we can observe that $S_6$ and $T_6$ can also
recover from the failures by using the equations from $\bfP_2$ and
$\bfP_3$. However, $S_5$ and $T_5$ cannot recover from the failure
since they can only obtain one equation from $\bfP_1$ in two
variables that corresponds to failures on ($S_2,T_2$) and
($S_5,T_5$). In Figure \ref{fig:bipartite}.(c), path $\bfP_2$ does
not exist, and ($S_6$,$T_6$) is protected only by path $\bfP_3$,
which protects two failed connections. Therefore, it cannot
recover from the failure. However, ($S_2$, $T_2$) can still
recover its data units by using path $\bfP_1$.

In general, this procedure needs to be performed for every
possible failure pattern that needs to be protected against, for
checking whether all nodes can still recover the data unit that
they are interested in. However, usually the set of failure
patterns to be protected against is the set of all single link
failures or more generally the set of all possible $M \geq 1$ link
failures. Those $M$ link failures can happen anywhere, on primary
paths or protection paths.


Next, we consider general conditions for data recovery. First, we
describe the general model for multiple failures. In order to make
expressions simple, we assume that the data unit obtained by a
node of a failed connection, say $S_i$, from protection path
$\bfP_k$ is the sum of the data units from $\bfS_k$, $\bfT_k$.
Adding up with $\alpha_{i\leftrightarrow j,k}d_i$, which is the
data units generated at node $S_i$, we denote this sum by $p_k$
where $p_k = y_{\sigma^{-1}(S_i) \rightarrow S_i} +
z_{\tau^{-1}(S_i) \rightarrow S_i} + \alpha_{i\leftrightarrow
j,k}d_i$. Note that $d_i$ is the local data units, which is always
available. In this case, each node on one protection path $P_k$
obtains the same equation in terms of the same variables. By
denoting the set of failed primary connections protected by
$\bfP_k$ as $F(\bfP_k)$, the equation for this protection path
$\bfP_k$ is
\begin{equation}
\label{eq:var}
\sum_{(S_i\leftrightarrow T_j) \in F(\bfP_k)}
\alpha_{i\leftrightarrow j,k} (d_i+u_j) = p_k.
\end{equation}

In equation \eqref{eq:var}, each $d_i+u_j$ is considered as one
variable and the coefficients assigned to $d_i$ and $u_j$ are the same. Each
node of a failed connection will obtain one equation from each
intact protection path that protects it and consequently forms a
system of linear equations. The number of equations that node $S_i$
obtains is the number of intact protection paths that protect
$S_i$. The number of variables is the total number of failed
connections protected by the protection paths that also provide
protection to the failed connection between $S_i$ and $T_j$. $S_i$
needs to solve the system of equations and obtain $d_i+u_j$. By
subtracting $d_i$, it can get $u_j$, which is the data unit $S_i$
wants to receive while $T_j$ can retrieve the data $d_i$ by
subtracting $u_j$ from $d_i+u_j$.

Each protection path maps to an
equation in terms of a number of variables representing the
combination of the data units generated at two end nodes of the
failed connections protected by this path. We can form a system of equations that consists of at most $K$
equations like equation \eqref{eq:var} where $K$ is the total
number of protection paths. Each failure of a primary path
introduces a variable whereas each failure occurring on a
protection path erases the corresponding equation from the
matrix. In general, the system of equations that a node
obtains also depends on the topology. If all of the connections
are not protected by the same protection paths, there are zeros in
the coefficient matrix because a failed connection is not
protected by all protection paths, implying that some variables
will not appear in all equations.

In order to recover from any failure pattern of $M$ failures, we
require the following necessary conditions.

\begin{theorem}
In order for the network to be guaranteed protection against any
$M$ link failures, the following necessary conditions should be
satisfied.
\begin{enumerate}
\item Each node should be protected by at least $M$ link-disjoint
protection paths.

\item Under any failure pattern with $M$ failures, a subset of
equations that each node obtains should have a unique solution.

\end{enumerate}
\end{theorem}
\emph{proof:} The first condition can be shown by contradiction.
If a node is protected by $M-1$ protection paths, the failure
could happen on these $M-1$ protection paths and on the primary
path in which this node participates. Then, this node does not
have any protection path to recover from its primary path failure.

The second condition is to ensure that each node can recover the
data unit under any failure pattern with $M$ failures. Note that
for necessary condition, we don't require that the whole system of
equations each node obtains has unique solution because one node
is only interested in recovering the data unit sent to it. As long
as it can solve a subset of the equations, it recovers from its
failure. \endproof
We emphasize that the structure of the equations depends heavily
on the network topology, the connections provisioned and the
protection paths. Therefore it is hard to state a more specific
result about the conditions under which protection is guaranteed.
However, under certain structured topologies it may be possible to
provide a characterization of the conditions that can be checked
without having to verify each possible system of equations. 

%
%
%

For example, if all connections are protected by $M$ protection
paths, it is easy to see the sufficient condition for data
recovery from any $M$ failures is that the coefficient matrix of
the system of equations each node obtains under any failure
pattern with $M$ failures has full rank. As will be shown next,
our coefficient assignment methods are such that the sufficient
conditions above hold.

Next we construct a $K\times N$ matrix to facilitate the
discussion of coefficient assignment. According to the encoding
protocol, each connection $S_i-T_j$ has coefficient
$\alpha_{i\leftrightarrow j,k}$ for encoding on $\bfP_k$. In
general, there are at most $K\times N$ coefficients for a network
with $N$ primary paths $S_{i_1}\leftrightarrow T_{j_1},
S_{i_2}\leftrightarrow T_{j_2},\ldots,S_{i_l}\leftrightarrow
T_{j_l},\ldots,S_{i_N}\leftrightarrow T_{j_N}$ and $K$ protection
paths $\bfP_1,\bfP_2,\ldots,\bfP_K$. We form a $K\times N$
matrix $\cal A$ where ${\cal A}_{kl}=\alpha_{i_l\leftrightarrow
j_l,k}$ if $S_{i_l}\leftrightarrow T_{i_l}$ is protected by
$\bfP_k$, ${\cal A}_{kl}=0$ otherwise. Here, $l$ is the index for
primary paths and each column of $\cal A$ corresponds to a primary
path. Each row of $\cal A$ corresponds to a protection path. This
matrix contains all encoding coefficients and some zeros induced
by the topology in general. It is easy to see that under any
failure pattern, the coefficient matrix of the system of equations
at any node of any failed connection is a submatrix of matrix
${\cal A}$. We require these submatrices of ${\cal A}$ to have
full rank. We shall discuss the construction of ${\cal A}$, i.e.,
assign proper coefficients in Section \ref{sec:coeff}.

\section{Encoding coefficient assignment}\label{sec:coeff}
In this section, we shall discuss encoding coefficient assignment
strategies for the proposed network coding schemes, i.e.,
construct ${\cal A}$ properly. Under certain assumptions on the
topology, two special matrix based assignments can provide tight
field size bound and efficient decoding algorithms. We shall also
introduce matrix completion method for general topologies.

Note that the coefficient assignment is done before the actual
transmission. Once the coefficients have been determined, during
data transmission they need not be changed. Thus, for the schemes
that guarantee successful recovery with high probability, we can
keep generating the matrix ${\cal A}$ until the full rank
condition discussed at the end of the previous section satisfies.
This only needs to be done once. After that, during the actual
transmission, the recovery is successful for sure.

\subsection{Special matrix based assignment}
In this and the next subsection, we assume that all primary paths are protected
by the same protection paths. This implies that matrix $\cal A$
only consists of encoding coefficients. It does not contain zeros
induced by the topology. Thus, we can let $\cal A$ to be a matrix
with some special structures such that any submatrix of $\cal A$
has full rank. The
network will be able to recover from any failure pattern with $M$
(or less) failures. Without loss of generality, we shall focus on
the case when $M=K$, where $K$ is the number of protection paths. If $M$ failures happen, in which $t_1$
failures happen on primary paths, each node will get
$M-(M-t_1)=t_1$ equations with $t_1$ unknowns corresponding to
$t_1$ primary path failures. The $t_1\times t_1$ coefficient
matrix is a square submatrix of $\cal A$ and they are the same for
each node under one failure pattern.

First, we shall show a Vandermonde matrix-based coefficient
assignment. It requires the field size to be $q\geq N$. If all
failures happen on primary paths, the recovery at each node is
guaranteed. In this
assignment strategy, we pick up $N$ distinct elements from
$GF(q)$: $\lambda_1,\ldots,\lambda_N$ and assign them to each
primary paths. At nodes $S_{i_l}$ and $T_{j_l}$, $\lambda_l^{k-1}$
is used as encoding coefficient on protection path $\mathbf{P}_k$,
i.e., ${\cal A}_{kl} = \alpha_{i_l\leftrightarrow
j_l,k}=\lambda_l^{k-1}$. In other words, $\cal A$ is a Vandermonde
matrix \cite[Section 6.1]{shulin}:
\begin{equation*}\label{eq:van}
\left[
\begin{array}{cccc}
1 & 1 & \cdots & 1 \\
\lambda_{1} & \lambda_{2} & \cdots & \lambda_{N} \\
\lambda_{1}^2  & \lambda_{2}^2 & \cdots & \lambda_{N}^2 \\
\cdots & \cdots & \cdots & \cdots \\
\lambda_{1}^{K-1}  & \lambda_{2}^{K-1} & \cdots & \lambda_{N}^{K-1}
\end{array}
\right].
\end{equation*}

Suppose $M$ failures happen on primary paths, the indices of
failed connections are $e_1,\ldots,e_M$, every node gets a system of linear equations with coefficient matrix having this form:
\begin{equation*}
\left[
\begin{array}{cccc}
1 & 1 & \cdots & 1 \\
\lambda_{e_1} & \lambda_{e_2} & \cdots & \lambda_{e_M} \\
\lambda_{e_1}^2  & \lambda_{e_2}^2 & \cdots & \lambda_{e_M}^2 \\
\cdots & \cdots & \cdots & \cdots \\
\lambda_{e_1}^{M-1}  & \lambda_{e_2}^{M-1} & \cdots &
\lambda_{e_M}^{M-1}
\end{array}
\right].
\end{equation*}
This matrix is a $M\times M$
Vandermonde matrix. As long as $\lambda_{e_1}, \lambda_{e_2},\ldots,
\lambda_{e_M}$ are distinct, this matrix is invertible and
$S_{i_{e_1}}$ can recover $u_{j_{e_1}}$. We choose
$\lambda_1,\ldots,\lambda_N$ to be distinct so that the submatrix
formed by any $M$ columns of $\cal A$ has full rank. The smallest
field size we need is the number of connections we want to
protect, i.e., $q \geq N$. Moreover, the complexity of solving
linear equation with Vandermonde coefficient matrix is $O(M^2)
$\cite{NumericalC}. Thus, we have a more efficient decoding
because if the coefficients are arbitrarily chosen, even if it is
solvable, the complexity of Gaussian elimination is $O(M^3)$.

If $M-t_1$ failures happen on protection paths, we require that
any $t_1\times t_1$ square submatrix formed by choosing any $t_1$
columns and $t_1$ rows from $\cal A$ has full rank. Although the
chance is large, the Vandermonde matrix can not guarantee this for
sure \cite[p.323,problem
7]{TheoECC},\cite{lacan04},\cite{shpar00}. We shall propose
another special matrix to guarantee that for combined failures,
the recovery is successful at the expense of a slightly larger
field size compared to Vandermonde matrix assignment.

In order to achieve this goal, we resort to Cauchy matrix \cite{TheoECC}, of
which any square submatrix has full rank if the entries are chosen
carefully.

\begin{definition}
Let
$\{x_1,\ldots,x_{m_1}\},\{y_1,\ldots,y_{m_2}\}$ be two sets of
elements in a field $F$ such that
\begin{itemize}
\item[(i)] $x_i + y_j \neq 0, \text{~ } \forall i\in
\{1,\ldots,m_1\} \text{~ } \forall j\in \{1,\ldots,m_2\}$;
\item[(ii)]$\forall i,j\in \{1,\ldots,m_1\},i\neq j: x_i\neq x_j$
and $\forall i,j\in \{1,\ldots,m_2\},i\neq j: y_i\neq y_j$.
\end{itemize}
The matrix $C=(c_{ij})$ where $c_{ij}= 1/(x_i+y_j)$ is called a
Cauchy matrix.
\end{definition}
If $m_1=m_2$, the Cauchy matrix becomes square and its determinant
is \cite{TheoECC}:
\begin{equation*}
\det(C) = \frac{\prod_{1\leq i<j\leq m_1}
(x_j-x_i)(y_j-y_i)}{\prod_{1\leq i,j\leq m_1} (x_i+y_j)}
\end{equation*}
Note that in $GF(q)$ where $q$ is some power of $2$, the addition
and subtraction are equivalent. Therefore, as long as
$x_1,\ldots,x_{m_1},y_1,\ldots,y_{m_1}$ are distinct, Cauchy
matrix has full rank and its any square submatrix is also a Cauchy
matrix (by definition) with full rank. For our protection problem,
we let matrix $\cal A$ to be a $K\times N$ Cauchy matrix.
$\{x_1, \ldots, x_K\},\{y_1, \ldots, y_N\}$ are chosen to be distinct.
Thus, the smallest field size we need is $K+N$. Suppose there are
$t_1$ failures on primary paths and $M-t_1$ failures on protection
paths, the coefficient matrix of the system of equations obtained
by a node is a $t_1\times t_1$ submatrix of $\cal A$. It is still
a Cauchy matrix by definition and invertible. Thus, the network
can be recovered from any $M$ failures. Moreover, the inversion
can be done in $O(t_1^2)$ \cite{blomer95xorbased}, which provides an
efficient decoding algorithm.

\subsection{Random assignment}
We could also choose the coefficients from a large finite field.
More specifically, we have the following claim \cite{CooperRanMat}.
\begin{claim}\label{clm:1err_ran}
When all coefficients are randomly, independently and uniformly
chosen from $GF(q)$, the probability that a $t_1$-by-$t_1$ matrix
has full rank is $p(t_1) = \Pi_{i=1}^{t_1}(1-1/q^i)$, $1\leq t_1\leq M$.
\end{claim}
Under one failure pattern with $t_1$ failures on the primary paths
and $M-t_1$ failures on the protection paths, every failed
connection obtains the equations that have the same $t_1$-by-$t_1$
coefficient matrix. The probability that it is full rank is
$p(t_1)$ and it goes to 1 when $q$ is large. Note that there are
$\sum_{t_1 = 1}^M {N \choose t_1}{M \choose {M-t_1}}$ possible
failure patterns when the total number of failures is $M$. Thus,
by union bound, the probability of successful recovery under any
failure pattern with $M$ failures is $1-\sum_{t_1 = 1}^M {N
\choose t_1}{M \choose {M-t_1}} (1-p(t_1))$, and it approaches 1
as $q$ increases.

\subsection{Matrix completion for general topology}
If the primary paths are protected by different protection paths,
like in Figure \ref{fig:3-paths-example}, there are some zeros in
$\cal A$ induced by the topology. We want to choose encoding
coefficients so that under every failure pattern with $M$
or less failures, the coefficient matrix of the system of
equations obtained by every node is invertible. We can view the
encoding coefficients in $\cal A$ as indeterminates to be decided.
The matrices we require to have full rank are a collection ${\cal
C}_{\cal A}$ of submatrices of $\cal A$, where ${\cal C}_{\cal A}$
depends on the failure patterns and the network topology. Each
matrix in ${\cal C}_{\cal A}$ consists of some indeterminates and
some zeros. The problem of choosing encoding coefficients can be
solved by matrix completion \cite{harvey05}. A simultaneous
max-rank completion of ${\cal C}_{\cal A}$ is an assignment of
values from $GF(q)$ to the indeterminates that preserves the rank
of all matrices in ${\cal C}_{\cal A}$. After completion, each
matrix will have the maximum possible rank. Matrix completion can
be done by deterministic algorithms \cite{harvey05}. Moreover,
simply choosing a completion at random from a sufficiently large
field can achieve the maximum rank with high probability
\cite{lovasz79}.  Hence, we can choose encoding coefficients
randomly from a large field.

\section{ILP Formulation for Single-link Failure}
\label{modified-formulation}

The problem of provisioning the working paths and their protection
paths in a random graph is a hard problem. This is due to the fact
that the problem of finding link disjoint paths between multiple
pairs of nodes in a graph is known to be NP-complete
\cite{Vygen95}. Therefore, in this section we formulate an integer
linear program that optimally provisions a set of unicast
connections, and their protection paths against single-link
failure. The optimality criterion is the minimization of the sum
of the working and protection resources.

The problem can be stated as follows:
{\it
Given an bidirectional graph $G=(V,E)$ and a traffic demand matrix
 of unicast connections, $\mathbb N$, establish a connection for each
 bidirectional traffic request $j \in \mathbb N$, and a number of protection
 paths that travel all the end nodes of the connections in $\mathbb N$,
 defined by set $\mathbb C$, such that:
\begin{itemize}
\item
A path protecting a connection must pass through the end nodes of the
 connection.
\item
The connections jointly protected by the same path must be
 mutually link disjoint, and also link disjoint from the protection
 path.
\item
The total number of edges used for both working and protection paths
 is minimum.
\end{itemize}
}
We also assume that the network is uncapacitated.

\begin{figure}[t]

\psfig{figure=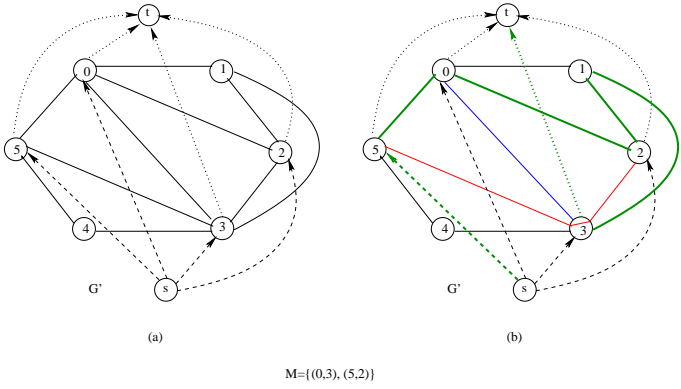,width=90mm} \caption{An example to show: (a)
the graph $G$ in solid line and its modified graph $G^\prime$;
(b) the provisioning of the connections ((0-3) and
  (5-3-2)) and their protection path (s-5-0-2-1-3-t), where the two
  links (s-5) and (3-t) are not included in the cost of the protection
  circuit.}
\label{fig:G}
\end{figure}

In order to formulate this problem, we modify the graph $G$ to obtain
the graph $G^\prime$ by adding a hypothetical source $s$ and a
hypothetical sink $t$. We also add a directed edge from $s$
to each node $v$, where $v\in \mathbb C$, as well as a directed edge from each
such node $v$ to $t$.
An example is shown in Figure \ref{fig:G}. Figure \ref{fig:G}.(a) shows a graph $G$ with six nodes and ten bidirectional edges and the corresponding modification to the graph $G^\prime$ given two traffic requests $\mathbb N=\{(0,3),(5,2)\}$.
Figure \ref{fig:G}.(b) shows the provisioning of the two
connections in $\mathbb N$ and their protection path from $s$ to $t$.
Therefore, the problem of finding the
protection paths turns out to be establishing connections from node
$s$ to $t$ that traverse all the nodes $v\in \mathbb C$. For each subset of connections that are protected together, the two ends nodes of these traffic requests have to be traversed by the same protection path.

This disjoint paths routing problem can be formulated with ILP as
follow: (Note that $G=(V,E)$ and $G^\prime=(V^\prime,E^\prime)$ denote
the original and modified graph in the formulation).
It is to be noted that the number of protection paths must satisfy:
\[
1 \leq \mathrm{number~of~protection~paths} \leq N.
\]

We may have more than one protection path because it is possible
that the primary connections are partitioned into several sets and
each set of primary connections share the protection of  path. However, the worst case is that each primary path requires a unique protection path (the case of 1+1 protection), which results in  a total of $N$ protection paths. In the formulation, therefore, we have a maximum of $2N$ paths:

\begin{itemize}
\item
Connections indexed from 1 to $N$ are the ones given by the set $\mathbb N$,
and these should be provisioned in the network.
\item
Connections indexed from $N{+}1$ to $2N$ are hypothetical
connections, which correspond to protection connections, and at least
one of them should be provisioned.
\end{itemize}

The ILP is formulated as a network flow problem, where there is a flow
of one unit between each pair of end nodes of a connection, and there
is also a flow of one unit from $s$ to $t$ for each protection path.

We define the following parameters, which are input to the ILP:

\begin{tabular}{l p{2.5in}}

$G(V,E)$: & the original network graph\\
$G^\prime (V^\prime, E^\prime)$: & the modified graph\\
$\mathbb N$: & the set of unicast connections\\
$c_{mn}$: & a constant, the cost of link $(m,n)\in E$\\
$v_j$: & set of end nodes of connection $j$ in $\mathbb N$, $v_j = \{s_j, t_j\}$, which are different notations from the previous definition of a connection, denoted by $S_i$, $T_j$ where $i,j$ are the indices for the nodes.

\end{tabular}

We also define the following binary variables which are computed by
the ILP:


\begin{tabular}{l p{2.8in}}
$f^{i}_{mn}$ & binary, equals 1 if the protection path $i$ traverses link $(m,n)$ in $G$\\
$Zf^{i}_m$ & integer, the number of times that the node $m\in V$ is traversed by path $i$\\
$U^{i}_j$ & binary, equals 1 if connection $j$ is protected by path $i$\\
$p^{j}_{mn}$\label{ILP:p} & binary, equals 1 if the working flow of $j$ traverses link $(m,n)\in G$\\
$q^{j}_{mn}$ & binary, equals 1 if the protection flow of $j$ traverses link $(m,n)\in G$\\
$Zp^{j}_m$ & integer, the number of times that node $m\in V$ is traversed by the working flow of $j$\\
$Zq^{j}_m$ & integer, the number of times that node $m\in V$ is traversed by the protection flow of $j$
\end{tabular}


{~}\\
\textbf{The objective function is:}
\[
\text{Minimize:}~~~ \sum_{(m,n)\in E}(\sum_{1\leq j\leq N}{p^{j}_{mn}c_{mn}}+
\sum_{N < i\leq 2N}{f^{i}_{mn}c_{mn}})
\]

The objective function minimizes the total cost of links used by the
working paths (first term) and by the protection paths (second term). Note that a protection path at $s$ and end at $t$ in the modified graph, $G^\prime$, but we only consider the cost of links in the original graph $G$.

{~}\\
\textbf{The constraints are such that:}

\begin{enumerate}
\item[1)]
\textbf{Working Flow Conservation:}

\begin{eqnarray}
&&\sum_{\{n:(s_j,n)\in E\}}{p^{j}_{s_jn}}=1,~~j\leq N ; \label{eqn:const1}\\
&&\sum_{\{n:(m,n)\in E\}}{p^{j}_{mn}=2Zp^{j}_m}, ~\forall m\in
V{\setminus} c_j. \label{eqn:const3}
\end{eqnarray}

The constraints \eqref{eqn:const1} 
and \eqref{eqn:const3} are standard flow conservation for working traffic which ensures
that a bidirectional path is established between end nodes $s_j$ and $t_j$ of connection $j$.

\item[2)]
\textbf{Protection Flow Conservation:}

\begin{eqnarray}
&&\textrm{For }\forall j\leq N, ~N < i\leq 2N: \nonumber\\
&&\sum_{\{n:(s_j,n)\in E\}}{q^{j}_{s_jn}}=1; \label{eqn:const4}\\
&&\sum_{\{n:(m,n)\in E\}}{q^{j}_{mn}=2Zq^{j}_m}, ~\forall m\in
V{\setminus} v_j; \label{eqn:const5}
\end{eqnarray}

Constraints \eqref{eqn:const4} and \eqref{eqn:const5} make sure that each connection $j$ has a protection flow.

\begin{eqnarray}
&&\sum_{\{n:(s,n)\in E^\prime\}}{f^{i}_{sn}}\leq 1;
\label{eqn:const6}\\
&&\sum_{\{n:(m,n)\in E\}}{f^{i}_{mn}=2Zf^{i}_m}, ~\forall m\in V; \label{eqn:const7}
\end{eqnarray}

The flow conservation of protection paths is ensured by constraints \eqref{eqn:const6} and \eqref{eqn:const7}. It is worth noting that not every protection path $i$ $(N{<}i{\leq} 2N)$ is required unless it is used for protection.

\begin{eqnarray}
&& \sum_{N < i \le 2N}{U^{i}_j} = 1; \label{eqn:const10}\\
&& \frac{1}{N}\sum_{j\le N}{U^{i}_j} \leq \sum_{\{n:(s,n)\in E^\prime\}}{f^{i}_{sn}}; \label{eqn:const11}\\
&& f^{i}_{mn} \geq q^j_{mn}+U^i_j -1, ~\forall (m,n)\in E; \label{eqn:const12}
\end{eqnarray}

Each working flow should be protected by exactly one protection path, guaranteed by constraint \eqref{eqn:const10}. Meanwhile, any protection path $i$ is provisioned only if it is used to protect any working path $j$. Otherwise, we do not need to provision it. Therefore, equation \eqref{eqn:const11} ensures this constraint. Furthermore, constraint \eqref{eqn:const12} ensures that if a protection path $i$ protects connection $j$, it should traverse the same links used by the protection flow $q^j_{mn}$.

\item[3)]
\textbf{Protection Path Sharing:}

\begin{eqnarray}
&&\textrm{For }\forall (m,n)\in E, ~N < i\leq 2N: \nonumber\\
&& p^j_{mn} + q^j_{mn} \leq 1, ~~~ \forall j\leq N; \label{eqn:const13}\\
&& p^j_{mn} + f^{i}_{mn} + U^i_j \leq 2,  ~~\forall j\leq N; \label{eqn:const14}\\
&& p^j_{mn} + p^k_{mn} + U^i_j + U^i_k \leq 3, ~\forall j<k\leq N. \label{eqn:const15}
\end{eqnarray}

The working flow and protection flow of each connection $j$ should be link disjoint, reflected by constraint \eqref{eqn:const13}. Each protection path may protect multiple connections so that it needs to traverse multiple corresponding protection flows. Thus, each protection path should also be link disjoint to all the working flow it protects. This constraint is ensured by equation \eqref{eqn:const14}. Meanwhile, if two connections are protected by the same path $f$, their working flow should also be link disjoint such that codewords can be decodes at each end nodes through the protection path. The last constraint is guaranteed by equation \eqref{eqn:const15}.

\end{enumerate}

The total number of variables used in the ILP is $(3N|V|+3N|E|+N^2)$ and the total number of constraints is $(6N|V|+2N + 2N^2|E| + N|E| + N^2(N-1)|E|)$, which is dominated by $O(N^3|E|)$.


\section{Numerical Results}
\label{sec:results}

\begin{figure}[htp]
\centerline{\psfig{figure=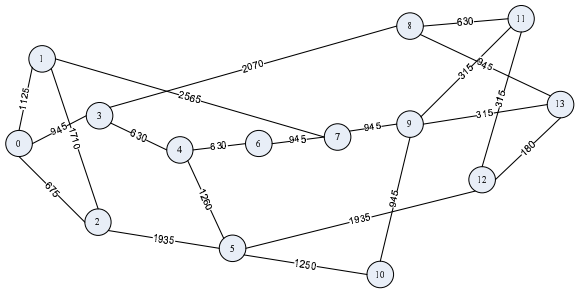,width=3.0in}}
\caption{NSFNET (N=14, E=21)}
\label{fig:NSF}
\end{figure}

\begin{figure}[htp]
\centerline{\psfig{figure=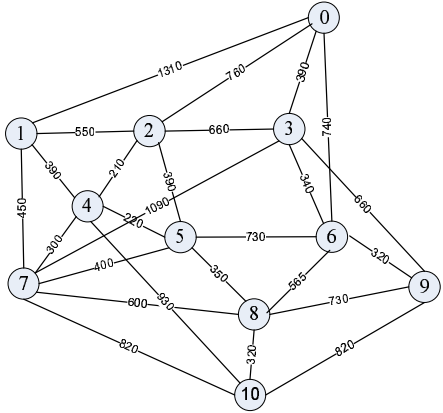,width=2.3in}}
\caption{COST239 (N=11, E=26)}
\label{fig:cost239}
\end{figure}

This section presents numerical results of the cost of our proposed protection scheme and compares it to 1+1 protection and Shared Backup Path Protection (SBPP) in terms of total resource requirements for protection against single-link failure. SBPP has been proven to be the most capacity efficient protection scheme and can achieve optimal solutions \cite{Rama99}. However, it is also a reactive protection mechanism and takes time to detect, localize and recover from failures. We consider two realistic network topologies, NSFNET and COST239, as shown in Fig. \ref{fig:NSF} and \ref{fig:cost239}, respectively. Both networks are bidirectional and each bidirectional span $e$ has a cost $c_e$, which equals the actual distance in kilometers between two end nodes.

\begin{figure}[htp]
\centerline{\psfig{figure=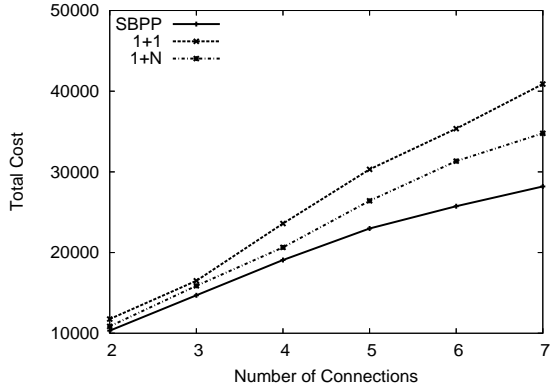,width=3.0in}}
\caption{Comparison of total cost in NSFNET}
\label{fig:NSF_cost}
\end{figure}

\begin{figure}[htp]
\centerline{\psfig{figure=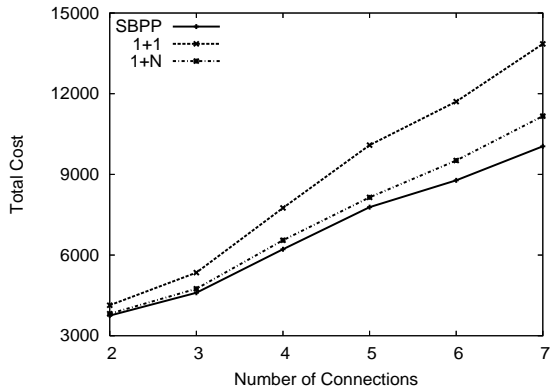,width=3.0in}}
\caption{Comparison of total cost in COST239 network}
\label{fig:cost239_cost}
\end{figure}

We first compare three schemes in terms of the total connection and protection provisioning cost in both networks as shown in Fig. \ref{fig:NSF_cost} and \ref{fig:cost239_cost}, respectively. We obtained the results by formulating the problems as ILPs using three different approaches. The x-axis denotes the number of connections in the static traffic matrix and y-axis denotes the total network design cost. Each value is the average cost over ten independent cases and all approaches used identical traffic requests for each case.

 Since SBPP is the most capacity efficient
scheme, it achieves the minimum cost. 1+N approach uses much lower
cost than 1+1, but is higher than SBPP in both networks. We
express the extra cost ratio of a scheme over SBPP by:
$(Cost_{scheme}-Cost_{SBPP})/Cost_{SBPP}$. The extra cost ratio of
1+N in NSFNET increases from 5.2\% to 23\% as the number of
connections increases from 2 to 7. Meanwhile, the extra cost ratio
of 1+1 over SBPP increases from 12\% to 45\%, which is almost
twice that of 1+N at each case. The advantage of 1+N over 1+1 in
COST239 is even more significant than NSFNET due to the larger
average nodal degree, 4.6, compared to, 3, in NSFNET. Hence, there
is a higher chance for multiple primary paths to share the same
protection path, which results in lower overall cost. Based on the
results, we can observe that the extra cost ratio of 1+N over SBPP
in COST239 increases from 1.8\% to 11.1\% whereas the ratio of 1+1
over SBPP increases from 10.2\% to 38\%, as the number of
connections increases from 2 to 7. Actually, the cost of using 1+N
is very close to the optimal in COST239 network. The extra cost
required by 1+N over the optimal solution is less than 27\% of
that achieved by 1+1 scheme.

In fact, if we only consider the cost of protection, i.e. exclude the cost of connection provisioning, 1+N protection uses much lower resources than 1+1 protection. For example, by examining one network scenario where there are seven connections in COST239 network, the average protection cost of using SBPP, 1+N and 1+1 protection schemes is 3586.0, 4313.5 and 6441.5, respectively. The saving ratio of 1+N to 1+1 is around 33\%, which is higher than the saving ratio of joint capacity cost (19.3\%). This example further illustrates the cost saving advantages of using 1+N protection over 1+1 protection.

In summary, 1+N protection has a traffic recovery speed which is comparable 1+1 protection. However, it performs significantly better than 1+1 scheme in terms of protection cost. Compared with the most capacity efficient protection scheme, SBPP, 1+N protection performs close to SBPP in terms of total capacity cost in dense networks. However, SBPP takes much longer to recover from failures due to the long switch reconfiguration time and traffic rerouting, which are not required in 1+N protection.

\section{Conclusions}
\label{sec:conclusions} This paper has introduced a resource
efficient, and a fast method for providing protection for a group
of connections such that a second copy of each data unit
transmitted on the working circuits can be recovered without the
detection of the failure, or rerouting data. This is done by
linearly combining the data units using the technique of network
coding, and transmitting these combinations on a shared set of
protection circuits in two opposite directions. The reduced number
of resources is due to the sharing of the protection circuit to
transmit linear combinations of data units from multiple sources.
The coding is the key to the instantaneous recovery of the
information. This provides protection against any single link
failure on any of the working circuits. The paper also generalized
this technique to provide protection against multiple link
failures.

The method introduced in this paper improves the
technique introduced in \cite{Kamal09} and \cite{Kamal07a}. In
particular, (a) it requires fewer protection resources, and (b) it
implements coding using a simpler synchronization strategy.
A cost comparison study of providing protection against single
link failures has shown that the proposed technique introduces a
significant saving over typical protection schemes, such as 1+1
protection, while achieving a comparable speed of recovery. The
numerical results also show that the cost of our 1+N scheme is
close to SBPP, the most capacity efficient protection scheme.
However, the proposed scheme in our paper provides much faster
recovery than SBPP.

\begin{biography}{Ahmed E. Kamal}
Ahmed E. Kamal (S'82-M'87-SM'91)is a professor of Electrical and Computer Engineering at Iowa State University.  His research interests include high-performance networks,
optical networks, wireless and sensor networks and performance evaluation.
He is a senior member of the IEEE, a senior member of the Association
of Computing Machinery, and a registered professional engineer.
He was the co-recipient of the 1993 IEE Hartree Premium for papers published in
Computers and Control in IEE Proceedings for his paper entitled
Study of the Behaviour of Hubnet, and the best paper award of the
IEEE Globecom 2008 Symposium on Ad Hoc and Sensors Networks Symposium.
He served on the technical program committees of numerous conferences and
workshops, was the organizer and co-chair of the first and second Workshops
on Traffic Grooming 2004 and 2005, respectively,
and was the chair of co-chair of the Technical Program Committees of a number of
conferences including the
Communications Services Research (CNSR) conference 2006,
the Optical Symposium of Broadnets 2006, and the Optical Networks and Systems
Symposium of the IEEE Globecom 2007, the 2008 ACS/IEEE International
Conference on Computer Systems and Applications (AICCSA-08), and
the ACM International Conference on Information Science, Technology
and Applications, 2009.
He is also the Technical Program co-chair of the Optical
Networks and Systems Symposium of the IEEE Globecom 2010.
He is on the editorial boards of the Computer Networks journal,
and the Journal of Communications.
\end{biography}
\vspace{-0.5in}
\begin{biography}{Aditya Ramamoorthy}
Aditya Ramamoorthy received his B. Tech degree in Electrical Engineering from the Indian Institute of Technology, Delhi in 1999 and the M.S. and Ph.D. degrees from the University of California, Los Angeles (UCLA) in 2002 and 2005 respectively. He was a systems engineer at Biomorphic VLSI Inc. till 2001. From 2005 to 2006 he was with the data storage signal processing group at Marvell Semiconductor Inc. Since Fall 2006 he has been an assistant professor in the ECE department at Iowa State University. He has interned at Microsoft Research in summer 2004 and has visited the ECE department at Georgia Tech, Atlanta in Spring 2005. His research interests are in the areas of network information theory and channel coding.
\end{biography}
\vspace{-0.5in}
\begin{biography}{Long Long}
Long Long received his B.Eng degree in Electronic Information
Engineering from Huazhong University of Science and Technology,
Wuhan, China in 2002 and M.Sc degree in Software Engineering from
Peking University, Beijing, China in 2005. Since fall 2006, he has
been a Ph.D student in ECE department of Iowa State University,
USA. His research interests are in the area of traffic grooming
and survivability of optical networks.
\end{biography}
\vspace{-0.5in}
\begin{biography}{Shizheng Li}
Shizheng Li received his B.Eng degree in information engineering
from Southeast University (Chien-Shiung Wu Honors College),
Nanjing, China, in 2007. He worked on error correction codes in
National Mobile Communications Laboratory at Southeast University
during 2006 and 2007. Since Fall 2007, he has been a Ph.D. student
in the Department of Electrical and Computer Engineering, Iowa
State University. His research interests include network coding,
distributed source coding and network resource allocation. He
received Microsoft Young Fellowship from Microsoft Research Asia
in 2006. He is a student member of IEEE.
\end{biography}


\begin{thebibliography}{10}

\bibitem{Zhou00}
D.~Zhou and S.~Subramaniam, ``Survivability in optical networks,'' {\em IEEE
  Network}, vol.~14, pp.~16--23, Nov./Dec. 2000.
\bibitem{LiR2010}
S.~Li and A.~Ramamoorthy, ``Protection against link errors and failures using network coding,''{\em IEEE Transactions on Communications}, to appear.
Available: http://arxiv.org/abs/0905.2248

\bibitem{Stamatelakis00b}
D.~Stamatelakis and W.~D. Grover, ``Ip layer restoration and network planning
  based on virtual protection cycles,'' {\em IEEE Journal on Selected Areas in
  Communications}, vol.~18, no.~10, pp.~1938--1949, 2000.

\bibitem{Grover04}
W.~D. Grover, {\em Mesh-based survivable networks : options and strategies for
  optical, MPLS, SONET, and ATM Networking}.
\newblock Upper Saddle River, NJ: Prentice-Hall, 2004.

\bibitem{Fragouli06}
C.~Fragouli, J.-Y. LeBoudec, and J.~Widmer, ``Network coding: An instant
  primer,'' {\em ACM Computer Communication Review}, vol.~36, pp.~63--68, Jan.
  2006.

\bibitem{Schupke03}
D.~Schupke and R.~Prinz, ``Performance of path protection and rerouting for wdm
  networks subject to dual failures,'' in {\em Optical Fiber Conference},
  pp.~209--210, 2003.

\bibitem{Kim03}
S.~Kim and S.~Lumetta, ``Evaluation of protection reconfiguration for multiple
  failures in WDM mesh networks,'' in {\em Optical Fiber Conference},
  pp.~210--211, 2003.

\bibitem{Zhang06}
J.~Zhang, K.~Zhu, and B.~Mukherjee, ``Backup provisioning to remedy the effect
  of multiple link failures in wdm mesh networks,'' {\em IEEE Journal on
  Selected Areas in Communications}, vol.~24, pp.~57--67, Aug. 2006.

\bibitem{Choi04}
H.~Choi, S.~Subramaniam, and H.-A. Choi, ``Loopback recovery from double-link
  failures in optical mesh networks,'' {\em IEEE/ACM Transactions on
  Networking}, vol.~12, pp.~1119--1130, Dec. 2004.

\bibitem{He05b}
W.~He, M.~Sridharan, and A.~K. Somani, ``Capacity optimization for surviving
  double-link failures in mesh-restorable optical networks,'' {\em Photonic
  Network Communications}, vol.~9, pp.~99--111, Jan. 2005.

\bibitem{Liu05}
Y.~Liu, D.~Tipper, and P.~Siripongwutikorn, ``Approximating optimal spare
  capacity allocation by successive survivable routing,'' {\em IEEE/ACM
  Transactions on Networking}, vol.~13, pp.~198--211, Feb. 2003.

\bibitem{Rama99}
S. Ramamurthy and B. Mukherjee, ``Survivable WDM mesh networks.
part I-protection,'' {\em in Proceedings of IEEE INFOCOM, 1999.}

\bibitem{Kamal06a}
A.~E. Kamal, ``1+N protection in optical mesh networks using network coding on
  p-cycles,'' in {\em the proceedings of the IEEE Globecom}, 2006.

\bibitem{Kamal07a}
A.~E. Kamal, ``1+N protection against multiple faults in mesh networks,'' in
  {\em the proceedings of the IEEE International Conference on
  Communications (ICC)}, 2007.

\bibitem{Kamal09}
A.~E. Kamal, ``1+N Network Protection for Mesh Networks:
Network Coding-Based Protection using p-Cycles,'' {\em IEEE/ACM Transactions on
  Networking}, Vol. 18, No. 1, Feb. 2010, pp. 67--80. 

\bibitem{Ahlswede00}
R.~Ahlswede, N.~Cai, S.-Y.~R. Li, and R.~W. Yeung, ``Network information
  flow,'' {\em IEEE Transactions on Information Theory}, vol.~46,
  pp.~1204--1216, July 2000.

\bibitem{Vygen95}
J. Vygen, "NP-completeness of some edge-disjoint paths problems",
{\em Discrete Appl. Math.}, vol.~46, pp.~83--90, 1995.

\bibitem{Bhandari99}
R. Bhandari,
{\em Survivable Networks: Algorithms for Diverse Routing}.
Springer, 1999.

\bibitem{NumericalC}
W.~H. Press, B.~P. Flannery, S.~A. Teukolsky, and W.~T. Vetterling,
  \emph{Numerical Recipes in C: The Art of Scientific Computing}, 2nd~ed.\hskip
  1em plus 0.5em minus 0.4em\relax Cambridge University Press, 1992.

\bibitem{TheoECC}
F.~J. MacWilliams and N.~J.~A. Sloane, \emph{The Theory of Error-Correcting
  Codes}.\hskip 1em plus 0.5em minus 0.4em\relax North Holland, 1977.

\bibitem{blomer95xorbased}
J.~Blomer, M.~Kalfane, R.~Karp, M.~Karpinski, M.~Luby, and D.~Zuckerman, ``An
  xor-based erasure-resilient coding scheme,'' Int. Comput. Sci. Inst.,
Berkeley, CA, TR-95-048, 1995. [Online]. Available:
 citeseer.ist.psu.edu/blomer95xorbased.html
\bibitem{lacan04}
J.Lacan and J.Fimes, ``Systematic MDS erasure codes based on Vandermonde
  matrices,'' \emph{IEEE Communications Letters}, vol. 8, no.9, Sep.2004.

\bibitem{shpar00}
I.~E. Shparlinski, ``On singularity of generalized Vandermonde matrices over
  finite fields'', {\em Finite Fields and Their Applications}, vol.~11, no.~2, pp.~193-199, 2005.
  
\bibitem{harvey05}
N.~J.~A. Harvey, D.~R. Karger, and K.~Murota, ``Deterministic network coding by
  matrix completion,'' in \emph{SODA '05: Proceedings of the sixteenth annual
  ACM-SIAM symposium on Discrete algorithms}, 2005, pp. 489--498.

\bibitem{lovasz79}
L.Lovasz, ``On determinants, matchings and random algorithms,'' in \emph{Fund.
  Comput. Theory 79, Berlin}, 1979.
  
\bibitem{shulin}
S.~Lin and D.~J. Costello, \emph{Error control coding: fundamentals and
  applications}.\hskip 1em plus 0.5em minus 0.4em\relax Prentice Hall, 2004.

\bibitem{CooperRanMat}
C.~Cooper, ``On the distribution of rank of a random matrix over a finite field,'' in \emph
{Random Struct. Algorithms}, vol. 17, no. 3-4, pp.197--212, 2000.


\end{thebibliography}
\end{document}